\documentclass[prd,aps,floats,epsfig,eqsecnum,nofootinbib]{revtex4}
\usepackage{amssymb,amsmath,epsfig,graphicx}
\newcommand{\be}{\begin{equation}}
\newcommand{\ee}{\end{equation}}
\newcommand{\bea}{\begin{eqnarray}}
\newcommand{\eea}{\end{eqnarray}}

\begin{document}
\title{\bf The quantum inflaton,
primordial perturbations and CMB fluctuations}
\author{\bf F. J. Cao$^{(a,b)}$}
\email{francao@fis.ucm.es}
\author{\bf H. J. de Vega$^{(c,a)}$}
\email{devega@lpthe.jussieu.fr}
\author{\bf N. G. S\'anchez$^{(a)}$}
\email{Norma.Sanchez@obspm.fr}
\affiliation{
 (a) Observatoire de Paris, LERMA.  Laboratoire Associ\'e au CNRS UMR 8112.
 \\   61, Avenue de l'Observatoire, 75014 Paris, France. \\
  (b) Grupo Interdisciplinar de Sistemas Complejos (GISC) and 
     Departamento de F\'{\i}sica At\'omica, Molecular y Nuclear. 
     Universidad Complutense de Madrid. E-28040 Madrid, Spain. \\
  (c) LPTHE, Universit\'e Pierre et Marie Curie (Paris VI) 
     et Denis Diderot (Paris VII),  Laboratoire Associ\'e au CNRS UMR 7589.
Tour 24, 5\`eme. \'etage, 4, Place Jussieu, 75252 Paris cedex 05, France.}
\begin{abstract}
We compute the primordial scalar, vector and tensor metric
perturbations arising from {\em quantum field} inflation. 
Quantum field  inflation takes into
account the {\em nonperturbative quantum} dynamics of the inflaton
consistently coupled to the dynamics of the (classical)
cosmological metric. For chaotic inflation, the quantum treatment
avoids the unnatural requirements of an initial state with all the energy 
in the zero mode. For new inflation it
allows a consistent treatment of the explosive particle production
due to spinodal instabilities. Quantum field inflation (under
conditions that are the quantum analog of slow roll) leads, upon
evolution, to the formation of a condensate starting a regime of
effective classical inflation. We compute the primordial
perturbations taking the dominant quantum effects into account.
The results for the scalar, vector and tensor primordial
perturbations are expressed in terms of the classical inflation
results. For a $N$-component field in a 
$O(N)$ symmetric  model, adiabatic fluctuations dominate while 
isocurvature or entropy fluctuations are negligible. The results agree
 with the current WMAP observations and predict corrections to  the power 
spectrum in classical inflation. Such  corrections are estimated
to be of the order of $ \frac{m^2}{N \; H^2} $, where $m$ is the inflaton 
mass and $H$ the Hubble constant at
the moment of horizon crossing. An upper estimate turns to be about $ 4\% $ 
for the cosmologically relevant scales.
This quantum field treatment of inflation
provides the foundations to the classical inflation and permits to compute
quantum corrections to it.
\end{abstract}
\date{\today}
\maketitle

\tableofcontents

\section{Introduction}

Inflation is a stage of accelerated expansion in the very early
Universe \cite{infbooks,revkolb}. The present observations make inflationary 
cosmology the leading theoretical framework to explain the homogeneity,
isotropy and flatness of the Universe, as well as the observed
features of the cosmic microwave background. The WMAP results 
\cite{wmap1,wmap2,wmap3} 
confirm the basic tenets of the inflationary paradigm.

There are many different models for inflation and most (if not all) of them 
invoke one or several scalar fields, the inflaton(s), whose dynamical evolution
coupled with the space-time geometry leads to an inflationary epoch. The
inflaton is a  scalar field  which provides an
effective description for  field condensates in the grand unified
theories (GUT). The inflaton field  is just an effective description of the 
particle dynamics and may not correspond to any real particle (even unstable). 
Fortunately, we do not need to know the detailed microscopical description 
given by the GUT to get the cosmological evolution. 
Indeed, a more precise description should be possible from a microscopic GUT.
Somehow, the inflaton is to the  microscopic GUT theory
like the Ginzburg-Landau theory of superconductivity is to the
microscopic BCS superconductivity theory. 

Most treatments of inflation studies the evolution of the inflaton
as a homogeneous {\bf classical} scalar field. The quantum field
theory interpretation is that this classical homogeneous field
configuration is the expectation value of a quantum field
operator in a translational invariant quantum state. In the
classical treatments, the evolution of this coherent field
configuration is studied through classical equations of motion,
while fluctuations of the scalar field around this classical value
are treated perturbatively and quantum mechanically, and provide
the seeds for the scalar density perturbations of the metric
\cite{infbooks}.

However, since the energy scale of inflation is so high (the GUT scale),
it is necessary a full {\bf quantum} field theory description for the 
matter.
Only such a quantum treatment permits a consistent description of
particle production and particle decays.

An important class of inflationary models, the `large field'
models \cite{revkolb}, produce inflation starting from a large
field amplitude configuration that rolls down the potential (for
example: chaotic inflation). Another important class of
inflationary models, the `small field' models \cite{revkolb},
produce inflation starting from a small field amplitude
configuration near the false vacuum of a spontaneously broken
symmetry potential (for example: new inflation). The inflaton
background dynamics for these models is usually studied in a
classical framework and in order to have a long inflationary
period it is necessary that the field rolls down very slowly: for
these models various conditions have been obtained which are
different realizations of what we will call here the {\em
classical slow roll condition}, \be
 \dot{\tilde{\varphi}}^2 \ll |m^2| \; \tilde{\varphi}^2 \; .
\ee
This condition guarantees that there is inflation ($ \ddot a > 0 $)
and that it lasts long enough.
($ \tilde{\varphi} $ is the classical inflaton
field, $ m $ its mass, and the dot denotes cosmic time
derivative.)

One of the {\em shortcomings} of the {\em classical} chaotic
inflationary scenarios is the need of an initial state with quite
unnatural restrictions. In classical inflation the energy is
dominated by the zero mode which leads the dynamics. In quantum
inflation all modes contribute to the energy and one can choose,
in particular, initial states with zero expectation value for the
inflaton. Therefore, the initial state may not break the $ \varphi
\to -\varphi $ symmetry of the potential, while this symmetry is
always broken in classical inflation. On the other hand, classical
inflation scenarios do not allow a consistent treatment of the
particle creation.

In order to overcome the restrictions on chaotic inflation and
provide a consistent description of new inflation, a {\em quantum}
treatment for the inflaton dynamics is needed \cite{tsuinf,qnew,qnewA,dashin}. 
This quantum treatment must be {\em nonperturbative} in order to
consistently include  the contribution of the excited modes since
the energy is proportional to the inverse of the coupling. The
nonperturbative method we use here is the large $N$ expansion.
Thus, we  consider a $ N $ component inflaton field with an $ O(N)
$ invariant interaction. The presence of the $ O(N) $ symmetry
simplifies the calculation of the large $N$ limit but is not a
conceptual restriction.

The goal of this article is to compute the primordial
perturbations produced in the {\bf quantum} inflation scenarios,
and show its relation with those produced by  classical
inflation.

We consider the case where the gravitational and
the inflaton backgrounds are homogeneous, and the metric background is
of the flat FRW type,
\be
  ds^2 = dt^2 - a^2(t)(dx^2+dy^2+dz^2)\;.
\ee

We present a quantum field framework for
inflation that takes into account the nonperturbative quantum dynamics of
the inflaton consistently coupled to the dynamics of the
(classical) metric. This generalized framework avoids the
shortcomings of the classical chaotic and new inflation scenarios.
It also clarifies {\em how and when} a classical effective scenario
emerges.

Quantum gravity corrections can be neglected during
inflation because the energy scale of inflation $ \sim m
\sim M_{GUT} \sim 10^{-6} M_{Planck} $. That is, quantum gravity
effects are at most $ \sim 10^{-6} $ and can be neglected in this context.

Quantum field inflation (under conditions that are
the quantum analogue of slow roll) leads, upon evolution, to the
formation of a {\bf condensate} starting a regime of effective
classical field inflation. That is, the $N$-component quantum
inflaton becomes an effective $N$-component classical inflaton,
which can be directly expressed in terms of an effective
\emph{single}-field classical inflation scenario.
The action structure, parameters (mass
and coupling) and initial conditions for the effective classical
field description are fixed by those of the underlying quantum
field inflation. This condensate description allows an easy
computation of the primordial perturbations that takes into
account the dominant quantum effects. That is, the effects of the
quantum nature of the inflaton background (which is absent in
classical inflation), and the effects due to the quantum nature of
the inflaton and metric perturbations. This condensate description
provides the primordial perturbation spectrum for quantum field
inflation in terms of the well-known classical inflation results.
We show that quantum inflation allows a consistent computation of
the background and of the primordial perturbations with results in
agreement with the observations.

For a $O(N)$ symmetric model, adiabatic fluctuations dominate while 
isocurvature and entropy fluctuations are negligible in agreement with 
the WMAP
observations. Therefore, the presence of a large symmetry in multi field 
models is supported by observations. Non-symmetric multi field models 
produce 
sizable isocurvature and entropy fluctuations.

Furthermore, the classical inflation scenario emerges as an
effective description of the post-condensate inflationary period both for
the {\em background} and for the {\em perturbations}. Therefore,
the quantum treatment of inflation provides the {\em foundations} for 
classical inflation.

The cosmologically relevant density fluctuations (i.e. for the CMB 
anisotropies) in quantum inflation exhibit  corrections
compared with classical inflation. We find as the main source of such 
corrections the quantum changes in the effective mass felt by the 
cosmological fluctuations. An estimate for such quantum  corrections
to the power spectrum yields 
$ {\cal O}\left( \frac{m^2}{N \; H^2}\right) $ at
the moment of horizon crossing which turns to be about $ 4\% $ for the
cosmologically relevant scales [and $ N = O(1) $].

\bigskip

This paper is organized as follows:
in section \ref{classinf}, we briefly present the single-field classical
inflation scenario. In section \ref{quantinf}, we present the results
in quantum field inflation dynamics for the  inflaton and the scalar
factor  (subsection \ref{backqfi}), and for the scalar,
vector and tensor perturbations (subsection \ref{pertqfi}).
Both quantum chaotic and quantum new inflation are treated.
The conclusions are developed in section \ref{conclu}.
Finally, two appendices are devoted to the computation
of scalar and tensor perturbations in multiple-field inflation.

\section{Classical inflation} \label{classinf}

Let us briefly present first the single-field classical
inflation scenario. The action for the {\em classical inflaton
dynamics} is
\be
  \tilde{S}_{cl} = \tilde{S}_{g} + \tilde{S}_{m}
  + \delta\tilde{S}_{g} + \delta\tilde{S}_{m}
\ee
where $\tilde{S}_{gr} + \tilde{S}_{m}$ describes the dynamics of
the background for the metric and the inflaton, respectively
and $\delta\tilde{S}_{gr} + \delta\tilde{S}_{m}$ describe
 their perturbations. In the classical framework only the
perturbations are quantized. We use the tilde, $ \tilde{} \; $, to
denote the quantities in classical inflation.

The gravitational action and its perturbation are
\be \label{gravact}
  \tilde{S}_{gr} + \delta\tilde{S}_{gr} = -\frac{1}{16\pi G}
    \int{\sqrt{-g}\;d^4x\,R}
\ee where $ G $ is the universal gravitational constant, and $ R $
is the Ricci scalar for the complete metric $ g_{\mu\nu} $.
By expanding $ g_{\mu\nu} $ in terms of the background FRW metric and
its perturbation, $ g_{\mu\nu} = g_{\mu\nu}^{(FRW)} + \delta
g_{\mu\nu} $, the $ \tilde{S}_{gr} $ terms (those which do not
contain $ \delta g_{\mu\nu} $) and the $ \delta\tilde{S}_{gr} $ terms
can be identified. These terms take account of the dynamics of the FRW
background and of the dynamics of the metric perturbations
respectively. (Detailed expressions can be found in
\cite{revmuk}.)

\bigskip

The classical matter action and its perturbation are
\bea
  \tilde{S}_{m} + \delta\tilde{S}_{m} &=& \int{\sqrt{-g}\;d^4x
    \left[\frac12\,\partial_{\alpha}\tilde{\chi}\,
    \partial^{\alpha}\tilde{\chi} - \tilde{V}(\tilde{\chi}) \right]}\\
  &=& \int{d^4x\,a^3(t)\, \left[\frac12(\dot{\tilde{\chi}})^2
    - \frac12\frac{(\nabla\tilde{\chi})^2}{a^2(t)}
    - \tilde{V}(\tilde{\chi}) \right]}\;.
\eea
We consider here a potential of the form
\be\label{poti}
  \tilde{V}(\tilde{\chi}) = \frac12\,\tilde m^2\,\tilde{\chi}^2
    + \frac{\tilde{\lambda}}{4} \; \tilde{\chi}^4
    + \frac{\tilde m^2}{4\tilde \lambda} \;  \frac{1-\alpha}{2}\;,
    \quad\quad \mbox{ with } \tilde \alpha \equiv \mbox{sign}(\tilde
    m^2) = \pm 1     \;,
\ee
where $ \tilde m^2 > 0 $ describes chaotic inflation, and $ \tilde m^2 < 0 $
describes new inflation. Expanding the inflaton field as $
\tilde{\chi} =
\tilde{\varphi}+\delta\tilde{\varphi} $, $ \tilde{S}_{m} $ stands
for the terms without $ \delta\tilde{\varphi} $ and takes account of
the field background dynamics,
while $ \delta\tilde{S}_{m} $ containing the  $ \delta\tilde{\varphi}
$ terms describes the field perturbations dynamics.

\bigskip

The initial state for chaotic inflation is a highly excited field state,
{\em i.e.}, a state with large $ |\tilde{\varphi}| $, while for new
inflation is a lowly excited state {\em i.e.}, with small $
|\tilde{\varphi}| $.

In order to have a long inflationary period, it is
necessary that the field rolls  down towards the minimum very
slowly. For these models
various conditions have been obtained that are different
realizations of what we  call here the {\em classical slow roll condition}:
\be
 \dot{\tilde{\varphi}}^2 \ll |m^2| \; \tilde{\varphi}^2,
\ee
this condition guarantees that there is inflation and that it lasts long
enough.

\bigskip

We will denote $ |\tilde{\delta}_k^{(S)}(\tilde
m^2,\tilde{\lambda})|^2 $ and $ |\tilde{\delta}_k^{(T)}(\tilde
m^2,\tilde{\lambda})|^2 $, the spectrum of primordial scalar and
tensor perturbations, respectively, for classical inflation. These
spectrums have been only computed in the literature \cite{infbooks,
revmuk, revriotto, turnerspectindex} for {\em classical}
inflation, but not for {\bf quantum} inflation
(see, however ref.\cite{qnewA}).

\section{Quantum field inflation}  \label{quantinf}

On the other hand, the action for {\em quantum field inflaton} is
\be
  S_{q} = \tilde{S}_{g} + S_{m} + \delta\tilde{S}_{g} + \delta S_{m}
\ee
where $ \tilde{S}_{g} + S_{m} $ describes the dynamics of the
background, and $ \delta\tilde{S}_{g} + \delta S_{m} $ that of
the perturbations. The important difference with classical
inflation is that the dynamics of the inflaton background
($ S_m $) is computed here in quantum field theory.

The gravitational terms have the same expressions as in the
classical inflaton dynamics [Eq. (\ref{gravact})].

In our treatment we consider semiclassical gravity: the geometry is
classical and the metric obeys the semiclassical Einstein equations
where the r. h. s. is the expectation value of the quantum energy
momentum tensor.
(Quantum gravity corrections are at most of order
$ \sim m / M_{Planck} \sim M_{GUT} / M_{Planck} \sim 10^{-6} $
and can be neglected.)

In order to implement a nonperturbative treatment, we consider
a $N$-component inflaton field $ \vec\chi $.
The matter action, besides of being quantum, is for a $ N $
component inflaton $ \vec\chi $ and displays  a $ 1/N $ factor in the $
(\vec\chi^2)^2 $ term in order to allow a consistent implementation
of the large $ N $ limit.
\bea \label{qftact}
  S_m + \delta S_{m} &=& \int{\sqrt{-g}\;d^4x
    \left[\frac12\,\partial_{\alpha}\vec{\chi}\,
    \partial^{\alpha}\vec{\chi} - V(\vec{\chi}) \right]}\\
  &=& \int{d^4x\,a^3(t)\, \left[\frac12(\dot{\vec{\chi}})^2
    - \frac12\frac{(\nabla\vec{\chi})^2}{a^2(t)}
    - V(\vec{\chi}) \right]}\;,
\eea
where $ \vec \chi = (\chi_1, \;\ldots, \;\chi_N) $
\be
  V(\vec{\chi}) = \frac12\,m^2\,\vec{\chi}^2
    + \frac{\lambda}{8N}\,\left( \vec{\chi}^2 \right)^2 +
    \frac{Nm^4}{2\lambda}\frac{1-\alpha}{2}\;, \label{potential}
    \quad \quad \mbox{ with } \alpha \equiv \mbox{sign}(m^2) = \pm 1
    \;,
\ee
For positive $m^2$ the $O(N)$ symmetry is unbroken while it is
spontaneously broken for $ m^2 < 0 $. The first case describes
chaotic inflation and the second one corresponds to new inflation.

The quantum field $ \vec \chi $ can be expanded as its expectation
value $ \langle \vec\chi(x) \rangle $ plus quantum contributions
which are in general large and cannot be linearized (except for $
k/a $ much larger than the effective mass). We therefore split the
quantum contribution $ \vec\chi - \langle \vec\chi(x) \rangle $
into large quantum contributions $ \vec\varphi(x) $ (or
background), plus small quantum contributions $
\delta\vec\varphi(x) $. Thus, we express the $N$-component quantum
scalar field $\vec{\chi}$ as
\be \label{Phidec}
  \vec{\chi}(x) = \langle \vec\chi(x) \rangle + \vec\varphi(x) +
  \delta\vec\varphi(x)  \; .
\ee
The dynamics of $ \delta\vec\varphi(x) $ can be then linearized,
and includes the cosmologically relevant fluctuations, that is those
which had exited the horizon during the last $N_e \simeq 60$ efolds of
inflation.

Without loss of generality, we can chose the '1'-axis in the
direction of the expectation value $ \langle \vec\chi(x) \rangle $
of the inflaton, and collectively denote by $ \vec \chi_{\bot} $
its $ N-1 $ perpendicular directions. That is,
\be
\vec{\chi}(x)= \left({\chi}_{\parallel}(x), \; \vec{\chi}_{\bot}(x) \right) 
    \quad  ,  \quad 
\langle \vec\chi(x) \rangle = (\sqrt{N} \; \varphi(t),\; \vec0)
\; ,
\ee
then, Eq. (\ref{Phidec}) reads
\be
\vec{\chi}(x) = \left(\sqrt{N} \;  \varphi(t) + \varphi_{\parallel}(x), \; 
\vec\varphi_{\bot}(x)   \right)\; + \;
\left(\delta\varphi_{\parallel}(x), \;  \delta\vec\varphi_{\bot}(x)\right)
\ee
$ \varphi(t) , \; \varphi_{\parallel}(x) $ and $
\vec\varphi_{\bot}(x) $ are the inflaton background contributions
which come from the quantum expectation value and from the quantum
fluctuations respectively; $ \delta\vec\varphi(x) $ is the
perturbation contribution. (The factor $ \sqrt{N} $ is made
explicit for convenience.)

After expanding Eq.(\ref{qftact}) using Eq.(\ref{Phidec}), $
S_{m} $ stands for the terms without $ \delta\vec{\varphi} $ and
describes the inflaton background dynamics, while $\delta S_{m} $
stands for the remaining terms which describe the inflaton
perturbation dynamics.

Both $ (\varphi_{\parallel}, \vec\varphi_{\bot}) $ and $
(\delta\varphi_{\parallel}, \delta\vec\varphi_{\bot}) $ represent
quantum fluctuations of the field around its expectation value,
and both can be expanded in Fourier modes. The field modes which
contribute to the observable primordial perturbations (those that
had exited the horizon during the last  $N_e \simeq 60$ efolds) are
part of the perturbation $ (\delta\varphi_{\parallel},
\delta\vec\varphi_{\bot}) $; while the field modes with larger
spatial scales are part of the background $ (\varphi_{\parallel},
\vec\varphi_{\bot}) $. In momentum space, let us call
$\Lambda $ the $k$-scale that separates the perturbation from the background.
Namely, $\Lambda $ must be smaller than 
the characteristic $k$-scale at which the modes
exited the horizon  $ N_e \simeq 60$ efolds before the end of
inflation and larger than the $k$-scales that dominate the background.

The mode expansions for the background inflaton 
$ (\varphi_{\parallel}, \vec\varphi_{\bot}) $ and the inflaton perturbations 
$ (\delta\varphi_{\parallel}, \delta\vec\varphi_{\bot}) $ are then
\bea
  \varphi_{\parallel}(\vec x,t) &=&
    \frac{1}{\sqrt{2}} \int_0^{\Lambda}
    \frac{d^3 k}{(2\pi)^3} \left[ b_{k} \; g_{k}(t)
    \;e^{i\vec k \cdot \vec x} + b^{\dagger}_{k} \;
    g^*_{k}(t) \; e^{-i\vec k \cdot \vec x}\right]
    \label{sigmamexp} \; , \\
  \vec{\varphi_{\bot}}(\vec x,t) &=&
    \frac{1}{\sqrt{2}} \int_0^{\Lambda}
    \frac{d^3 k}{(2\pi)^3} \left[ \vec{a}_{k} \; f_{k}(t)
    \;e^{i\vec k \cdot \vec x} + \vec{a}^{\dagger}_{k} \;
    f^*_{k}(t) \; e^{-i\vec k \cdot \vec x}\right]
    \label{pimexp} \; , \\
  \delta\varphi_{\parallel}(\vec x,t) &=&
    \frac{1}{\sqrt{2}} \int_{\Lambda}^\infty
    \frac{d^3 k}{(2\pi)^3} \left[b_{k} \; g_{k}(t)
    \;e^{i\vec k \cdot \vec x} + b^{\dagger}_{k} \;
    g^*_{k}(t) \; e^{-i\vec k \cdot \vec x}\right]
    \label{dsigmamexp} \; , \\
  \delta\vec\varphi_{\bot}(\vec x,t) &=& \frac{1}{\sqrt{2}}
    \int_{\Lambda}^\infty
    \frac{d^3 k}{(2\pi)^3} \left[\vec{a}_{k} \; f_{k}(t)
    \;e^{i\vec k \cdot \vec x} + \vec{a}^{\dagger}_{k} \;
    f^*_{k}(t) \; e^{-i\vec k \cdot \vec x}\right] \label{dpimexp} \; ,
\eea 
with $ b_k $, $ \vec{a}_{k} $ and $ b^{\dagger}_k $, $
\vec{a}^{\dagger}_{k} $ being annihilation and creation operators,
respectively, satisfying the canonical commutation relations. The
background $ ( \varphi_{\parallel}, \vec{\varphi_{\bot}} 
) $, includes the modes with $ k < \Lambda $, while the
perturbations $ ( \delta\varphi_{\parallel},
\delta\vec{\varphi_{\bot}} ) $ include the modes with $ k >
\Lambda $.

For asymptotic values of $ k $ (hence $ k > k_{Planck} $) the modes
tend to the vacuum modes, ensuring the finiteness of the total energy.
The scale $ \Lambda $ is well above the $k$-modes that dominate
the bulk of the energy, and well below the cosmologically relevant modes.
The results are independent of the precise value of $ \Lambda $.

This is due to the fact that modes with $ k \gg m $ cannot be
significantly excited since the energy density of the universe
during inflation must be of the order $ \gtrsim 10 \; m^2 \;
M_{Planck}^2 $.

On the other hand, relevant modes for the large scale structure
and the CMB are today in the range from $ 0.1$ Mpc to $10^3$ Mpc.
These scales at the beginning of inflation correspond to physical
wavenumbers in the range
$$
e^{N_T-60} \; 10^{16} \, GeV < k < e^{N_T-60} \; 10^{20} \, GeV
$$
where $N_T$ stands for the total number of efolds (see for example
Ref. \cite{sd}).

Therefore, there is an intermediate $k$-range of  modes
which are neither relevant for the background nor for the observed
perturbation. $\Lambda$ is inside this $k$-range, and the results
are independent of its particular value. In usual cases we
 can safely choose for $ \Lambda $,
$$
10 \; m \lesssim \Lambda \lesssim 10^3 \; e^{N_T-60} \; m \;.
$$

\subsection{Quantum field inflation dynamics}  \label{backqfi}

We now describe the main features of the {\em background
dynamics}, {\em i.e.}, the $ a $, $ \varphi $ and $ (
\varphi_{\parallel}, \vec{\varphi_{\bot}} ) $ dynamics. We treat
the inflaton as a full quantum field, and we study its dynamics in
a selfconsistent classical space-time metric (consistent with
inflation at a scale well below the Planck energy density). The
dynamics of the space-time metric is determined by the
semiclassical Einstein equations, where the source term is given by
the expectation value of the energy momentum tensor of the quantum
inflaton field $ G_{\mu\nu} = 8\pi m_{Pl}^{-2}\langle T_{\mu\nu}
\rangle $. Hence we solve  {\em self-consistently} the coupled
evolution equations for the classical metric and the quantum
inflaton field.

The amplitude of the quantum fluctuations for a set of modes can
be large (in quantum chaotic inflation due to the initial state, and in
new inflation due to spinodal instabilities). This implies the
need of a non-perturbative treatment of the evolution of the
quantum state, and therefore we use the large $N$ limit method.

In the large $ N $ limit, the longitudinal quantum contributions $
\varphi_{\parallel} $ are subleading by a factor $1/N$ \cite{tsuinf,
 tsuboya, tsunos}.
Thus, the evolution equations for the inflaton background
at leading order in large $ N $
can be expressed in terms of its expectation value, $ \varphi(t) $, and the
mode functions $ f_k(t) $ of the transversal quantum contributions
$ \vec\varphi_{\bot} $.
In the large $N$ limit, the evolution equations for the inflaton background
are,
\begin{eqnarray}
&&\ddot\varphi + 3\,H\,\dot\varphi + {\cal M}^2\,\varphi = 0
  \label{eqNexpect} \\ 
&&\ddot f_k +3\,H\,\dot f_k + \left(\frac{k^2}{a^2} + {\cal
    M}^2\right)\,f_k = 0  \label{eqNmodes} \\
&&\mbox{with }\quad {\cal M}^2 = m^2 + \frac{\lambda}{2}\,\varphi^2 +
    \frac{\lambda}{2}\,\int_R{\frac{d^3k}{2(2\pi)^3}\,|f_k|^2}  \; ,
\label{masaef}
\end{eqnarray}
and for the scale factor ($ H \equiv \dot a / a $),
\be
H^2 = \frac{8\pi}{3\,m_{Pl}^2} \; \rho \; ;
\quad \quad \frac{\rho}{N} = \frac12 \, \dot\varphi^2 + \frac{{\cal
M}^4 - m^4}{2\lambda} + \frac{m^4}{2\lambda}\frac{1-\alpha}{2} +
\frac14 \int_R \frac{d^3k}{(2\pi)^3} \left(|\dot f_k|^2 +
\frac{k^2}{a^2}|f_k|^2\right)\;. \label{epsilon}
\ee
where $ \rho = \langle T^{00} \rangle $ is the energy density. The
pressure ($ p\, \delta_i^{\;j} = \langle T_i^{\;j} \rangle $) is
given by
\be
\frac{p+\rho}{N} = \dot\varphi^2 + \frac12 \int_R
\frac{d^3k}{(2\pi)^3} \left(|\dot f_k|^2 +
\frac{k^2}{3a^2}|f_k|^2\right)\;. \label{pressure}
\ee
The index $ R $ denotes the renormalized expressions of these
integrals \cite{tsuinf}. This means that we must subtract the appropriate 
asymptotic ultraviolet behaviour in order to make convergent 
the integrals in Eqs.(\ref{masaef})-(\ref{pressure}):
\begin{eqnarray}
&&|f_k|^2 \buildrel{k^2\gg B(t)}\over= \frac{1}{k \;
a^2(t)}\left[ 1- \frac{B(t)}{2k^2} + \frac{1}{8\; k^4}\left\{3 \;
B^2(t) + a(t) \frac{d}{dt}\left[a(t)\dot B(t)\right] \right\} +
{\cal O}\left(\frac{B^3(t)}{k^6}\right) \right]
\; ,\label{renosubs} \\
&&|\dot f_k|^2\buildrel{k^2\gg B(t)}\over=
\frac{k}{a^4(t)}\left\{ 1 + \frac{1}{2\,k^2}
\left[ B(t) +2\; \dot{a}^2(t) \right]
- \frac{1}{8\; k^4}\left[  B^2(t) + a^2(t) \ddot B(t) - 3 a(t)
\dot a(t) \dot B(t) + 4 \dot a(t) B \right]
+ {\cal O}\left(\frac{B^3(t)}{k^6}\right) \right\}\; , \nonumber
\end{eqnarray}
where
$$
B(t) = a^2(t) \left[{\cal M}^2(t) - \frac{{\cal R}(t)}{6} \right]
$$
and the scalar curvature is
$$
{\cal R}(t) = 6 \left[\frac{\ddot a(t)}{a(t)} + \frac{\dot
a^2(t)}{a^2(t)} \right] \;.
$$
Equations (\ref{eqNexpect})-(\ref{eqNmodes}) for the expectation value and
for the field modes are analogous to damped oscillator equations,
and the inflationary period ($ \ddot a > 0 $)  corresponds to the
overdamped regime of these damped oscillators.

\bigskip

We consider here two typical classes of quantum inflation models:
\begin{itemize}
\item
{\em (i) Quantum chaotic inflation}, where inflation is produced
by the dynamical {\em quantum} evolution of a excited initial pure
state with large energy density (more details and the
generalization to mixed states can be found in \cite{tsuinf}).
This state is formed by a distribution of excited modes. It can be
shown that the initial conditions for a general pure state are
given by fixing the complex values of $ f_k(0) $ and $ \dot f_k(0)
$. Among these four real (two complex) numbers for each $ k $
mode, one is an arbitrary global phase, and another is fixed by
the wronskian. The two remaining degrees of freedom fix the
occupation number for each mode and the relative phase between $
f_k(0) $ and $ \dot f_k(0) $. The coherence between different $ k
$ modes turns out to be determined by such relative phases.

\item
{\em (ii) Quantum new inflation}, where inflation is produced by
the dynamical {\em quantum} evolution of a state with small
inflaton expectation value, and small occupation numbers for the
quantum modes, evolving with a spontaneously broken symmetry
potential. (More details can be found in \cite{qnew,qnewA}.)

\end{itemize}

The two classes of quantum inflation models have important
differences in their initial state and in their background and
perturbation dynamics ({\em e.g.}, spinodal instabilities are
present in new inflation and not in chaotic inflation). However,
we stress here the common features which allow a unified treatment
of the computation for the primordial perturbations generated in
these models.

In this quantum field inflation framework we have found the
following \emph{generalized slow roll condition}
\be \label{gsrc}
\dot\varphi^2+\int_R{\frac{d^3k}{2(2\pi)^3}\,|\dot f_k|^2} \;\ll\; m^2
\left( \varphi^2 + \int_R{\frac{d^3k}{2(2\pi)^3}\,|f_k|^2} \right)
\ee
which guarantees inflation ($\ddot a > 0$) and that it lasts
long (for both scenarios). (This condition includes the classical
slow roll condition $ \dot\varphi^2 \ll m^2 \; \varphi^2 $ as a
particular case.) 
There is a wide class of quantum initial conditions satisfying
Eq.(\ref{gsrc}) and leading to inflation that lasts long enough \cite{tsuinf}.

The quantum field dynamics considered here leads to \emph{two
inflationary epochs}, separated by a condensate formation:

\begin{enumerate}

\item \emph{The pre-condensate epoch:} During this epoch the term
$ D \equiv \int \frac{d^3k}{(2\pi)^3} \; \frac{k^2}{a^2} \; |f_k|^2 $ in
Eq. (\ref{epsilon}) has an important contribution to the energy
density while it fastly decreases due to the exponential redshift
of the excitations ($ k/a \to 0 $). This epoch ends at a time
$\tau_A $ when the $ D $ contribution to the energy density
becomes negligible, {\em i.e.}, the $ k^2/a^2 $ contribution in
the background evolution equations is negligible at $ \tau = \tau_A $.

After outward horizon crossing, the time dependence of the modes
factorizes and becomes $ k $ independent. The $ k^2/a^2 $ term in
Eq. \eqref{eqNmodes} becomes negligible, and all the modes satisfy
the same damped oscillator equation. For $ m^2 > 0 $ the modes
decrease (due to the damping), while for $ m^2 < 0 $ they grow
(due to spinodal instabilities). At the end of this epoch ($ t =
\tau_A $) all the relevant modes for the background dynamics have
exited the horizon, and the time dependence factorization allows
to consider them as a zero mode condensate.

\item \emph{The post-condensate  quasi-de Sitter epoch}.
The enormous redshift of the previous epoch assembles the quanta into
a zero mode condensate, $ \tilde{\varphi}_{eff} $, given by\cite{qnew,qnewA}
\bea
\tilde{\varphi}_{eff}^1(t) &=& \sqrt{N} \; \varphi(t) \cr
\tilde{\varphi}_{eff}^i(t) &=&
\sqrt{ \int{\frac{d^3k}{2(2\pi)^3}\,|f_k(t)|^2}} \quad \mbox{ for }i = 2,
\ldots, N
\eea
with constant direction in the field space [due to the $ O(N) $
invariance of the potential], and modulus
\be
\tilde{\varphi}_{eff}(t) = \sqrt{N}
\sqrt{\varphi^2(t)+\int{\frac{d^3k}{2(2\pi)^3}\,|f_k(t)|^2}}\;.
\ee
The modulus $ \tilde{\varphi}_{eff} $ verifies the classical
equations of motion,
\bea
&&\ddot{\tilde{\varphi}}_{eff} + 3 \,H \,\dot{\tilde{\varphi}}_{eff}
  + \tilde m^2 \; \tilde{\varphi}_{eff} + \tilde{\lambda}\;
  \tilde{\varphi}_{eff}^3 = 0\;, \label{eqclasphi} \\
&&H^2 = \frac{8 \pi}{3 m_{Pl}^2}\; \rho\;, \quad \quad
 \rho = \frac12\; \dot{\tilde{\varphi}}_{eff}^2
 + \frac12\; \tilde m^2 \; \tilde{\varphi}_{eff}^2 +
 \frac{\tilde{\lambda}}{4}\;\tilde{\varphi}_{eff}^4 \;, \label{epsilonclas} \\
&&\mbox{with } \quad  \tilde{\lambda} = \frac{\lambda}{2 N} \;,
  \quad \tilde m^2 = m^2 \;, \quad
  \tilde \alpha \equiv \mbox{sign}(\tilde m^2) = \pm 1 \;.
\eea
The pressure is given by
\be
p + \rho = \dot{\tilde{\varphi}}_{eff}^2 \;.
\ee
Therefore, the background evolution in this period can be {\em
effectively} described by a {\em classical} scalar field obeying
the evolution equation (\ref{eqclasphi}) and with initial
conditions defined at $ t = \tau_A $. Moreover, it is important to
stress that the {\em initial conditions for $
\tilde{\varphi}_{eff} $ are fixed by the quantum state}:
\be \label{iniclasphi}
\tilde{\varphi}_{eff}(\tau_A) = \sqrt{N}
\sqrt{\varphi^2(\tau_A)+\int{\frac{d^3k}{2(2\pi)^3}\,|f_k(\tau_A)|^2}}\;,
\ee
\end{enumerate}
Also the value of $ \tau_A $ depends on the full quantum evolution
{\bf before} the formation of the condensate. $ \tau_A $ is therefore a
function of the coupling, the mass and the quantum initial conditions
\cite{tsuinf}.

The previous result shows that after the formation of the
condensate (both for chaotic and for new inflation), the
background dynamics can be described by an effective classical
background inflation whose action structure, parameters
(mass and coupling) and initial conditions are fixed by those
of the underlying quantum field inflation.

Let us call $ N(t) $ the efolds remaining at time $ t $ till the end of 
inflation:
\be \label{nefo}
N(t) \equiv
\log\left[\frac{a_{end}}{a(t)}\right]
=\int_{t}^{t_{end}} dt' \; H(t') \; .
\ee
In particular, the total number of efolds is given by $ N_T =
\log\left[\frac{a_{end}}{a_{initial}}\right] $ which must be larger than $ N_e 
\simeq 60 $.

One particular consequence of quantum inflation, is that it can change 
the total number of e-folds $N_T$.
In chaotic inflation $N_T$ {\em decreases} if the initial state had excited modes with
non-zero wavenumber (for constant initial energy). For example, if the initial
energy is concentrated in a shell of wavenumber $ k_0 $ and for
simplicity the quadratic term dominates the potential Eq.(\ref{potential}), 
we have \cite{tsuinf}
\be \label{nefq}
N_T \simeq \frac{4\pi}{m_{Pl}^2\,m^2}\;
\frac{\rho_0}{\;1+(k_0/m)^2\;}
\ee
(where the classical result is recovered at $ k_0 = 0 $). We have
shown \cite{tsuinf} that there are enough efolds even for $k_0 \sim
80 \, m $ for reasonable choices of the initial energy density ($
\rho_0 = 10^{-2} m_{Pl}^4 $) and of the parameters (for instance, 
$ N \; m^2 /[ \lambda \;  m_{Pl}^2 ] = 2 \cdot 10^5 $).

\subsection{Primordial perturbations in quantum field inflation}
\label{pertqfi}

The relevant primordial perturbations are those that exited the
horizon during the last efolds of inflation \cite{infbooks}. As we
have seen in the previous subsection the background ($ a $, $
\varphi $, $ \varphi_{\parallel} $ and $ \vec{\varphi_{\bot}} $)
dynamics during the last efolds in the quantum field inflation
scenarios (both chaotic and new) are effectively classical. This
will allow to compute the relevant primordial perturbations for
these scenarios and express them in terms of the known perturbations for the
corresponding single-field classical scenarios.

The  more general metric perturbation $\delta g_{\mu\nu} $ can be
decomposed as usual in scalar $\delta g^{(S)}_{\mu\nu} $, vector
$\delta g^{(V)}_{\mu\nu} $ and tensor $\delta g^{(T)}_{\mu\nu} $
components \cite{revmuk, multiftent}
\be
 g_{\mu\nu} = g_{\mu\nu}^{(FRW)} + \delta g_{\mu\nu}
  = g_{\mu\nu}^{(FRW)} + \delta g^{(S)}_{\mu\nu} + \delta g^{(V)}_{\mu\nu}
  + \delta g^{(T)}_{\mu\nu} \;,
\ee
with
\bea
g_{\mu\nu}^{(FRW)} &=& a^2({\cal T}) \left(
  \begin{array}{cc}
    1               &   0  \\
    0               &   -\delta_{ij}
  \end{array} \right)   \\
\delta g^{(S)}_{\mu\nu} &=& a^2({\cal T}) \left(
  \begin{array}{cc}
    2\Phi           &   - \partial_i B  \\
    - \partial_i B  &   2[\Psi\delta_{ij}-\partial_i\partial_j E]
  \end{array} \right)   \label{dgS} \\
\delta g^{(V)}_{\mu\nu} &=& - a^2({\cal T}) \left(
  \begin{array}{cc}
    0               &   - S_i   \\
    - S_i           &   \partial_i F_j + \partial_j F_i
  \end{array} \right)   \label{dgV} \\
\delta g^{(T)}_{\mu\nu} &=& - a^2({\cal T}) \left(
  \begin{array}{cc}
    0               &   0  \\
    0               &   h_{ij}
  \end{array} \right)   \label{dgT} \; .
\eea
Here, ${\cal T}$ is the conformal time,  and we define
$$
{\cal H} \equiv \frac{1}{a} \frac{d a}{d{\cal T}}   = a \, H \quad , \quad
{\cal T} \equiv \int^t \frac{d\overline t}{a(\overline t)} \; \; ,
$$
$ \Psi , \;  \Phi , \; E ,  \; B , \;  S_i ,  \; F_i $ and $
h_{ij} $ are functions of space and time. $ \Psi ,  \; \Phi , \; E
$ and $ B $ are scalars, $ S_i $ and $ F_i $ are three-vectors and
$ h_{ij} $ is a three-tensor. $ i, j $ are three-spatial indexes
raised and lowered with $ \delta^{ij} $ and its inverse $
\delta_{ij} $, respectively. The following constraints are
imposed,
$$
 \partial^i S_i = \partial^i F_i = 0 \quad , \quad  h^i_i = 0 \quad , \quad
 \partial^j h_{ij} = 0 \quad ,
$$
in order to guarantee that $ S_i $ and $ F_i $
do not contain pieces that transform as scalars, and that $ h_{ij} $ do
not contain pieces that transform as scalars or vectors.
The gauge independence of the physical results allows us to choose
the longitudinal gauge ($ E = 0 ,  \; B = 0 $) for the scalar
perturbations, and the vector gauge ($ F_i = 0 $) for the vector
perturbations \cite{revmuk, multiftent}. The advantage of these
gauges is that the equations have the same form as with the gauge
invariant quantities
\be
\Phi^{(gi)} \equiv \Phi + \frac{1}{a} \; 
\frac{d}{d{\cal T}}[(B-\frac{d E}{d{\cal T}})a] \; ; \quad
\Psi^{(gi)} \equiv \Psi - {\cal H} \; (B-\frac{d E}{d{\cal T}}) \;; \quad
S_i^{(gi)}  \equiv S_i -\frac{dF_i }{d{\cal T}} \;.
\ee
On the other hand, as the space-space perturbations of the
energy-momentum tensor for the inflaton satisfies $ \delta T^i_j
\propto \delta^i_j $ we have $ \Psi = \Phi $.

\bigskip

The perturbations are usually described in terms of the following
spectral quantities
\bea
|\delta_k^{(S)}|^2 &\equiv& \frac{2k^3}{9\pi^2} \; \langle
\Phi_k^2 \rangle \;; \\
|\delta_k^{(V)}|^2 &\equiv& \frac{2 k^3}{9\pi^2} \; \langle
S_i(k)\, S^i(k) \rangle \;; \\
|\delta_k^{(T)}|^2 &\equiv& \frac{2 k^3}{9\pi^2} \;  \langle
 h_{ij}(k)\, h^{ij}(k) \rangle \;;
\eea
which are respectively the spectra for scalar, vector and
tensor perturbations. The index $ k $ in $ \Phi_k \equiv \Phi(k)
$, $ S_{i}(k) $, $ h^{ij}(k) $, \ldots denotes the $ k $ Fourier
component of the respective perturbation, defined as \be
 h^{ij}(x) = \int \frac{d^3 k}{(2\pi)^{3/2}} \;  \left[ h^{ij}(k)
 \, e^{i \vec k \cdot \vec x} +
 h^{ij*}(k) \, e^{-i \vec k \cdot \vec x} \right] \;.
\ee
The scalar and tensor spectral indexes, $ n_S $
and $ n_T $ and their respective runnings $ \frac{dn_S}{d ln k}  , \;
\frac{dn_T}{d ln k} $, are defined in the environment of a momentum
scale $ k_0 $ by 
\bea \label{spindexdef}
|\delta_k^{(S)}|^2 &\equiv& |\delta_{k_0}^{(S)}|^2
\left(\frac{k}{k_0}\right)^{n_S-1+\frac12 ( \frac{dn_S}{d ln
    k})\ln\left(\frac{k}{k_0} \right) } \quad , \cr \cr
|\delta_k^{(T)}|^2 &\equiv& |\delta_{k_0}^{(T)}|^2
\left( \frac{k}{k_0} \right)^{n_T+\frac12 \left( \frac{dn_T}{d ln
    k}\right) \ln \left( \frac{k}{k_0} \right)} \; . 
\eea
In the linear approximation, scalar, vector and tensor
perturbations evolve independently and thus can be considered
separately \cite{revmuk}.

\subsubsection{Scalar perturbations}

We now compute the {\em scalar metric perturbations} to the background,
these are tightly coupled to the inflaton perturbations, and therefore,
both perturbations have to be studied together \cite{revmuk,revriotto}.

We have shown that the background dynamics for quantum field
inflation can be separated in two epochs: before and after the
formation of the condensate.
As the first one is short, in the more natural scenarios
the last $ N_e \simeq 60 $ efolds take place after the formation of the
condensate. 
Thus, the cosmologically relevant scales of the perturbations
exit the horizon when the condensate was already formed.

Therefore, the dynamics of the perturbations after the formation
of the condensate is well approximated by that given by the
effective classical inflation background,
Eqs.(\ref{eqclasphi})-(\ref{iniclasphi}). Recall that the form of
the effective classical evolution, and the values of the
parameters and of the initial conditions for the condensate are
determined by the underlying quantum field theory through the
relations shown in the previous subsection \ref{backqfi}.

The pre-condensate period determines the initial conditions of the
cosmological relevant modes for the post-condensate epoch at time $
t = \tau_A $. These initial conditions are different from the vacuum
ones. We choose in general,
\be \label{condin}
f_k(0)=\frac{1}{\sqrt{\Omega_k}} \quad , \quad
{\dot f}_k(0)=- [i \Omega_k + H(0)+ \beta_k \; \omega_k(0)]f_k(0)
\; ,
\ee
where $ \Omega_k $ and $\beta_k$ are functions of $k$ that characterize
the initial state and $ \omega_k(t) = \sqrt{k^2 + {\cal M}^2(t)}$.
The initial modes for vacuum initial conditions (as those customary used in 
classical inflation) are
$$
f^{cl \; inf}_k(0)=\frac{1}{\sqrt{\omega_k(0)}} \quad , \quad
{\dot f}^{cl \; inf}_k(0)=- [i \omega_k(0) + H(0) ]f^{cl \; inf}_k(0)
\; ,
$$
However, the finiteness of the energy density [Eq.
(\ref{epsilon})] imposes that the difference between these initial
conditions and the vacuum ones must asymptotically vanish for 
$ k \to \infty $.
We see from Eqs.(\ref{epsilon}) and (\ref{renosubs}) that the mode functions
$|f_k(0)|^2$ and $|f^{cl \; inf}_k(0)|^2$ can differ asymptotically at 
most as
\be\label{delcoi}
{\cal O}\left(\frac{a^5(t) \; m^5}{k^5}\right) \; ,
\ee
where we used the inflaton mass as scale for the effective inflaton mass.

A further source of quantum inflation effects arises from the
evolution equations (\ref{eqNmodes}). The effective mass squared ${\cal M}^2 $
differs from the one used in classical inflation due to the last term in 
Eq. (\ref{masaef}) which contains the integral over the square modulus of the 
mode functions. The contribution of the modes $ k < \Lambda $ is taken into 
account
during the post-condensate epoch by  $ \tilde{\varphi}_{eff} $. The modes
$ k > \Lambda $ contribute to $  {\cal M}^2(t) $ as
\be\label{delL}
 \delta{\cal M}^2_{\Lambda}(t) \equiv
(\mbox{contribution to} \; {\cal M}^2(t) \; \mbox{from modes} 
\; k > \Lambda)
 \; \sim \lambda \; a^2(t) \; m^4 \; \int_{\Lambda}^{\infty}  
\frac{d^3k}{k^5} 
\sim   \lambda \; \frac{a^2(t) \; m^4}{\Lambda^2}
\ee
Furthermore, contributions from higher orders in $1/N$ will take the form,
\be\label{delN}
 \delta{\cal M}^2_{N}(t) \equiv
(\mbox{contribution to} \; {\cal M}^2(t) \; \mbox{from}\; \frac1{N} \;
\mbox{corrections}) \; \sim \frac{m^2}{N} \; .
\ee
Contributions $\delta{\cal M}^2(t)$ to the effective mass squared 
induce corrections in the square
modulus of the mode functions $|f_k(t)|^2$ of the order 
$ O([a^2 \; \delta{\cal M}^2]/k^3) $ as one sees from their large-$k$ 
behaviour Eq.(\ref{renosubs}).
As a result of  these three quantum effects:
initial conditions Eq.(\ref{delcoi}),   $ \delta{\cal M}^2_{\Lambda}(t) $ in
Eq.(\ref{delL}) and   $\delta {\cal M}^2_{N}(t) $ in Eq.(\ref{delN}), 
the mode functions result,
\be \label{f2}
\frac{ |f_k(t)|^2} { |f^{cl \; inf}_k(t)|^2} \buildrel{k \gg a \; m
  }\over= 1 +  {\cal O}\left(\frac{a^5(t) \; m^5}{k^5} \right) +
{\cal O}\left(\frac{a^2(t) }{k^2} \; 
\delta {\cal M}_{\Lambda}^2(t) \right) + {\cal O}\left(\frac{a^2(t)}{k^2} \; 
\delta {\cal M}_{N}^2(t) \right) \; ,
\ee
Using the above estimates for $ \delta {\cal M}^2_{\Lambda}(t) $ and  
$\delta {\cal M}^2_N(t) $ Eqs.(\ref{delL})-(\ref{delN})
and expressing $k$ in terms of the time where the mode exited the horizon 
through $ k \sim a \; H $  we obtain,
\be \label{f3}
\frac{ |f_k(t)|^2} { |f^{cl \; inf}_k(t)|^2} \buildrel{H \gtrsim m
}\over= 1 + {\cal O}\left(\frac{m^5}{H^5} \right)
+ {\cal O}\left(\frac{ \lambda \; a^2(t) \; m^4}{H^2 \;
\Lambda^2} \right) + {\cal O}\left(\frac{m^2}{N \; H^2} \right) \; .
\ee
The ratio $ \frac{m}{H} $ is time dependent and depends on the inflationary 
scenario considered. In chaotic inflation  $ \frac{m}{H} $ decreases with time
and takes a value  $ \frac{m}{H} \sim \frac15 $ when cosmologically relevant
scales cross the horizon, that is about 50 efolds before the end of inflation. 
Hence, we can make an  order of magnitude estimate
using the asymptotic approximation Eq. (\ref{f2}) at horizon crossing.
Explicit calculations have to be
performed in order to obtain the time dependence of these corrections.
We can however state that the corrections in Eq.(\ref{f3}) are multiplied
by functions of time [which are $ {\cal O}\left(1\right)$]. 

The third and fourth terms in  Eq.(\ref{f2})  and  (\ref{f3}) give
$k$-dependent corrections when we replace the time dependence present in 
both terms using  $ k \sim a(t) \;  H(t) $ at horizon crossing.

\bigskip

Since  $ \frac{m}{H} \sim \frac15 $, the second term in Eq.(\ref{f3}) 
yields corrections of the order
$ \sim 10^{-4} $ and can be neglected. 
The third term gives very small corrections for times  about 50 efolds 
before the end of inflation of the order $ \sim 10^{-9} \; 
\frac{m^2}{\Lambda^2} $ where we used that $ \lambda \sim 10^{-12} $
and $ \Lambda \gg m$.  The validity of the small $k$-large $k$ 
decomposition [Eqs.~(\ref{Phidec})-(\ref{dpimexp})] is confirmed by the
smallness of this third contribution, and by the smallness of
the correction induced by $ \delta{\cal M}^2_{\Lambda}(t) 
$ [Eq.(\ref{delL})] in the metric background through
the Friedmann equations.

In conclusion, we find that the quantum inflaton corrections 
to the power spectrum are of the order
\be \label{efecto}
\frac{m^2}{N \; H^2} \sim 4 \%  \; .
\ee
This value corresponds to $ N = O(1) $ and gives an upper estimate to 
the corrections.

Moreover, the corrections to the effective mass squared  given by 
Eqs. (\ref{delL})-(\ref{delN}) induce changes in the 
metric background through the Friedmann equation (\ref{epsilon}). 
Such changes on $ H(t) $ and $ a(t) $ produce changes of the order 
$ \frac1{N} $ in the mode functions. Such effects are larger
than those in Eqs. (\ref{f3}) and (\ref{efecto}). 
However, they could be similar to those appearing in classical inflation
since they arise from a change of the metric background.

Indeed, detailed calculations are needed to obtain a precise evaluation
of the quantum inflation effects on density fluctuations.

\bigskip

It must be noticed that a complete study of the cosmologically
relevant modes $ f_k(t) \; ,  ( k > \Lambda ) $ must include the graviton
in a gauge invariant way. The order of magnitude of the estimates given above
[Eqs. (\ref{f2})-(\ref{f3})] is not changed by that effect.

During the condensate period further quantum effects can also come
from the quantum interaction between the quantum inflaton
background and the quantum perturbations. However, these effects
are expected to be diluted due to the large difference in $ k $.

In addition, the pre-condensate period enters in the computation of the 
cosmologically relevant perturbations through the
determination of the initial state for the condensate at $ t =
\tau_A $.

\bigskip

In summary, we compute the scalar perturbations dynamics using the
effective classical inflaton background and considering initial
quantum vacuum conditions for the modes responsible of the
formation of cosmic structure. As we have shown this treatment
takes account of the dominant quantum effects: the quantum nature
of the inflaton background (absent in classical inflation), and
the quantum nature of the inflaton and metric perturbations.

The most general scalar density perturbation is the sum of an
adiabatic and an isocurvature (or entropy) perturbation
\cite{multiftent, revriotto}.
\be
\Phi_{k} = \Phi_{k\; ad} + \Phi_{k\; iso} \;.
\ee
The fluctuations in the different field components generate
entropy or isocurvature perturbations which is characteristic of 
multi field models.

A generic multi field model has all three contributions: adiabatic, 
isocurvature and mixture of them, all of the same
order. However, if the potential is completely symmetric for $ O(N) $
rotations in the internal space, the slow roll trajectories are
straight lines in field space. We show below that in this case
the isocurvature density perturbations are negligible and the adiabatic 
contributions dominate. The adiabatic density perturbations are then the same 
as in the {\it single-field case}. 

It is obvious that straight trajectories exists for $O(N)$ invariant
interactions, that is, one can always assume a solution moving in a
fixed direction in internal space. However, non-straight trajectories
can also exist for $O(N)$ invariant interactions. For such
trajectories at least two components of the field should be
nonzero and therefore, the `isospin' tensor $
\chi^a \; \Pi_{\chi}^b - \chi^b \; \Pi_{\chi}^a $
will be non-zero. We have restricted ourselves to $O(N)$ states with zero
`isospin' here as well as in refs.\cite{tsuinf,qnew,qnewA} since the
universe as a whole should be expected to be in a $O(N)$ invariant state.

Adiabatic perturbations are produced by the field fluctuations
parallel to the background inflaton trajectory in phase-space, and
have non-zero total energy density. On the other
hand, isocurvature perturbations are related to the field
perturbations in other directions (thus, they require a
multicomponent inflaton), and have vanishing total energy density.

In our case, due to the $ O(N) $ invariance of the potential,
the background inflaton solution does not change its direction in field space.
In this case, we show in Appendix A that
the \emph{isocurvature density perturbations} are \emph{negligible}
\bea
\Phi_{k\; iso} &=& 0 \; , \\
\mbox{implying} \qquad  \qquad 
| \delta_{k\; iso}^{(S)}(m^2,\lambda)|^2 &\equiv&
  \frac{2 k^3}{9\pi^2} \;  \langle \Phi_{k\; iso}^2 \rangle = 0 \\
| \delta_{k\; mix}^{(S)}(m^2,\lambda)|^2 &\equiv&
  \frac{2 k^3}{9\pi^2} \;  (\langle \Phi_{k\; iso} \Phi_{k\; ad}
  \rangle + \langle \Phi_{k\; ad} \Phi_{k\; iso}
  \rangle ) = 0 \;.
\eea
The symmetric multi field models (as here invariant under $ O(N) $)
are consistent with the last CMB data from WMAP \cite{wmap1,wmap2,wmap3}, 
as they indicate
that the adiabatic contribution dominates and give an upper bound
for isocurvature contributions. Initial conditions are consistent with being 
purely adiabatic.

It must be noticed that {\em non-symmetric} models with 
different masses or couplings for the different components of the
field, would lead to {\em non negligible} isocurvature density
perturbations, analogously to the classical case.

As a consequence of the straight trajectory of the background
inflaton in field space (see Appendix A), the power spectrum of
{\em adiabatic scalar perturbations} for the quantum field
inflation is
\be \label{scalpert}
| \delta_{k\;ad}^{(S)}(m^2,\lambda)|^2 =
\left| \tilde\delta_{k\;ad}^{(S)}\left(m^2,\;\frac{\lambda}{2 N} \right)
   \right|^2
\ee
where $ |\tilde\delta_k^{(S)}(m^2,\tilde{\lambda})|^2 $ is the power
spectrum of scalar perturbations for the {\bf single-field} classical
background inflation [Eqs.(\ref{eqclasphi})-(\ref{iniclasphi})].

The result Eq.(\ref{scalpert}) express the scalar
density perturbations for the  quantum field inflation in terms of
the associated effective classical background inflation, whose
action structure, parameters and initial conditions are determined
by  the underlying quantum field inflation. The relation 
[Eq.(\ref{scalpert})] allows to link to the primordial perturbations
from  classical field inflation. In terms of the slow roll parameters,
\bea\label{epseta}
&&\epsilon \equiv 2 \; M_{Pl}^2 \; \left(\frac{H'}{H}\right)^2
 = \frac12 \,  M_{Pl}^2 \; \left( \frac{V'}{V} \right)^2 \quad , \cr \cr
&& \eta \equiv  2 \; M_{Pl}^2 \; \frac{H''}{H}\quad ,  
\quad \eta_V \equiv
 M_{Pl}^2 \;  \frac{V''}{V} =  \eta + \epsilon \quad , \cr \cr
&& \xi \equiv 4 \; M_{Pl}^4 \;  \frac{H' \; H'''}{H^2} \quad ,  \quad
\xi_V \equiv   M_P^4 \; \frac{V' \; V'''}{V^2} = \xi + 3 \; \epsilon
\; \eta \quad ,
\eea
the adiabatic scalar perturbations can be expressed as
\bea \label{scalpertsr}
|\tilde{\delta}_{k\;ad}^{(S)}(m^2,\tilde{\lambda})|^2  &=&
  \frac{1}{8 \; \pi^2 \; M_{Pl}^2}
  \frac{H_{\cal H}^2}{\epsilon_{\cal H}}
  \left[ 1-2 \; \epsilon_{\cal H} + 2 (2-\gamma-\ln 2) \,
(2\epsilon_{\cal H} + \eta_{\cal H}) + 
{\cal O} \left(\epsilon_{\cal H}^2,
\eta_{\cal H}^2,\eta_{\cal H}\epsilon_{\cal H}\right)
  \right]  \cr \cr
&=& \frac{1}{12 \, \pi^2 \;  M_{Pl}^6} \; \frac{V^3}{V'^2} \left\{ 1 -
    \frac56 \;  M_{Pl}^2 \;  \left(\frac{V'}{V}\right)^2 + 
M_{Pl}^2\; (2-\gamma-\ln 2) \,\left[ \left(\frac{V'}{V}\right)^2 + 2 \; 
 \frac{V''}{V}\right] \right. \cr \cr
&+& \left. {\cal O} \left(  \left[ M_{Pl} 
\;\frac{V'}{V}\right]^4 \right) \right\}\; .
\eea 
with $ \gamma = 0.57721\ldots $ the Euler constant. All quantities are 
evaluated at the time of horizon crossing, when  
$ {\cal H} = k , \; i. e. \;  H \, a = k $.
This is stressed by the subscript $ {\cal H} $ .
Here, $  M_{Pl}^2 = \frac{1}{8 \, \pi \, G} = \frac{m_{Pl}^2}{8 \,
  \pi} $ and primes denote derivatives with respect to the inflaton
field $ \tilde{\varphi}_{eff} $ which satisfies,
\bea
&&\dot{\tilde{\varphi}}_{eff} = - 2 \; M_{Pl}^2 \;
H'(\tilde{\varphi}_{eff})  \quad , \cr \cr
&& H^2 \left( 1 - \frac{\epsilon}{3} \right) = \frac{1}{3 \;
  M_{Pl}^2} V(\tilde{\varphi}_{eff}) \; .
\eea
Using the relation Eq.(\ref{scalpert}) also allows to express
the scalar spectral index and its running [Eq.\eqref{spindexdef}] in
terms of the slow roll parameters as
\bea\label{ns1}
&&n_S = 1 - 4 \;  \epsilon_{\cal H} + 2 \; 
\eta_{\cal H}=1-6 \; \epsilon_{\cal
  H} + 2 \; \eta_V  \; , \cr \cr
&&\frac{d n_S}{d \ln k} = 10 \; \epsilon_{\cal H} \;  \eta_{\cal H}- 8
\; \epsilon_{\cal H}^2 - 2  \; \xi_{\cal H} = 16  \; \epsilon_{\cal H} \;
\eta_V  - 24  \; \epsilon_{\cal H}^2 - 2 \; \xi_V  \; ,
\eea
or in terms of the effective classical potential as
\bea \label{ns}
&&n_S = 1 - 3 \; M_{Pl}^2 \; \left( \frac{V'}{V} \right)^2 + 2  \;
  M_{Pl}^2 \;  \frac{V''}{V} \; , \cr \cr
&& \frac{d n_S}{d \ln k} = - 2 \; M_{Pl}^4 \; \frac{V' \;
V'''}{V^2} - 6  \; M_{Pl}^4 \;  \left( \frac{V'}{V} \right)^4 
+ 8  \; M_{Pl}^4 \;\frac{V'' \; V'^2 }{V^3} \; .
\eea
taken at the value of the field when the scale of interest
exited the horizon.

Expressions (\ref{scalpertsr}), (\ref{ns1}) and (\ref{ns}) get in 
addition quantum inflaton corrections of the order $ 4 \% $ 
[see Eq. (\ref{efecto})]. 

\medskip

For example, in chaotic inflation when $ \tilde\varphi \ll 
m / \sqrt{\lambda}, $ the potential [Eq.\eqref{epsilonclas}]
is dominated by the quadratic term, while for  $ \tilde\varphi \gg
m /\sqrt{\lambda} $ the quartic term dominates. Thus, for these
limiting cases the potential has the form
\be \label{potb}
V = \beta \; \tilde\varphi^b \; ,
\ee
(where $ \beta $ and $ b $ are positive constants) implying that
at $ N_e $ efolds before the end of inflation $ \tilde\varphi_{N_e} $
has the value 
\be \label{camf}
\tilde\varphi_{N_e}^2 \simeq 2 \; N_e \; b \;  M_{Pl}^2 \; .
\ee
where we have used Eq.(\ref{nefo}), i. e. $ N_e = \frac{1}{M_{Pl}^2} 
\int_{\tilde\varphi_{end}}^{\tilde\varphi} \frac{V}{V'} \; d\tilde\varphi
$ with $ \tilde\varphi_{end} \ll \tilde\varphi $.

Our convention for the amplitude of scalar perturbations $ |
\delta_{k\;ad}^{(S)}(m^2,\lambda)|^2  $ 
Eqs.(\ref{scalpert})-(\ref{scalpertsr}) is the same as the one
used by the WMAP collaboration \cite{wmap1,wmap2,wmap3} (called 
$ \Delta_{\cal R}^2(k) $ by them, WMAP only uses the leading order). 
The quoted $WMAP_{ext}$ + 2dFGRS +
Lyman $\alpha$ data  \cite{wmap1,wmap2,wmap3} for the overall primordial 
spectrum amplitude is  
$$ 
A(k_0=0.002\ {\rm Mpc}^{-1}) = 0.75^{+0.08}_{-0.09} (68\% CL).
$$
or $ 0.71^{+0.10}_{-0.11} (68\% CL) $ using  WMAP data alone. 
$ A(k) $ is related to $ | \delta_{k\;ad}^{(S)}(m^2,\lambda)|^2  $ by 
$$
 | \delta_{k\;ad}^{(S)}(m^2,\lambda)|^2 \equiv \Delta_{\cal R}^2(k) = 800
 \; \pi^2 \left(\frac53\right)^2 \frac{1}{T^2_{CMB}} \; A(k) \simeq
 2.95 \times 10^{-9} \;  A(k) \; .
$$
where $ T_{CMB} = 2.725 K $.
This factor comes from the relation between the CMB multipole
coefficients ($C_l$) and $   | \delta_{k\;ad}^{(S)}(m^2,\lambda)|^2
$. They are connected in the conventions of the  CMBFAST code used by
WMAP as 
$$
C_l = 4 \pi \; T^2_{CMB} \; \int \frac{9}{25} \;  |
\delta_{k\;ad}^{(S)}(m^2,\lambda)|^2 \; \left[ g_l(k) \right]^2 \; \frac{dk}{k}
$$
with $  g_l(k) $ being the radiation transfer function.

Using Eq.\eqref{scalpertsr}, this implies,
\be\label{v3vp}
\frac{V^3}{V'^2} = \frac{\beta}{b^2} \; \tilde\varphi^{b+2} 
\simeq 2.62 \cdot 10^{-7} \; M_{Pl}^6 \;,
\ee
for the scale when $ k_0 = 0.002\ {\rm Mpc}^{-1} $ exited the horizon.

For example, when the potential is dominated by the quadratic term,
Eq.(\ref{v3vp}) implies 
\be
m = 6.02 \cdot 10^{-6} M_{Pl} \simeq 1.45 \cdot 10^{13} GeV \quad , \quad 
\varphi_{N_e} = 15.5 \;  M_{Pl} \simeq 3.72  \cdot 10^{19} GeV \;, 
\ee
while if the potential is dominated by the quartic term, 
Eq.(\ref{v3vp}) implies
\be
\lambda =3.34 \cdot 10^{-13} \, N  \quad , \quad 
\varphi_{N_e} = 21.9 \;  M_{Pl} \simeq 5.26  \cdot 10^{19} GeV
\;.
\ee
Here the potentials have been evaluated at the field value Eq.\eqref{camf} 
for $ N_e = 60 $ efolds and  $ M_{Pl} = \frac{m_{Pl}}{\sqrt{8 \,
  \pi}} = 2.4 \cdot 10^{18} $ GeV. 

\bigskip

Plugging Eqs.(\ref{potb})-(\ref{camf}) in Eqs.(\ref{ns}), we obtain
\be
n_S = 1-\frac{b+2}{2 \, N_e} \quad , \quad \frac{d n_S}{d \ln k} =
-\frac{1}{2N_e^2}\left( b + 2 \right) \;.
\ee
Evaluating these expressions $ N_e = 60 $ efolds before the end of
inflation (which correspond to a typical scale
of astrophysical interest), if the quadratic term dominates we obtain
\be
n_S = 0.97 \; , \quad\quad \frac{d n_S}{d \ln k} = -5.5\cdot10^{-4} \; ;
\ee
while if the quartic term dominates we have
\be
n_S = 0.95 \; , \quad\quad \frac{d n_S}{d \ln k} = -8.3\cdot10^{-4} \; ;
\ee
The values obtained from these
examples are compatible with the current WMAP data
\cite{wmap1,wmap2,wmap3}. 
An exact scale-invariant spectrum (i.e. $ n_S = 1 , \; 
\frac{d n_S}{d \ln k} = 0 $) is not yet excluded at more than 
$2 \, \sigma $ level by WMAP data.

\subsubsection{Vector perturbations}

The vector metric perturbations [Eq. (\ref{dgV})] do not have any source
in their evolution equation, because the energy-momentum tensor for
a scalar field does not lead to any vector perturbation.
The $ (0i) $ components of the Einstein equations, in the absence of vector
perturbation sources, gives
\be
 \Delta S_i = 0 \;,
\ee
implying that there cannot be any space-dependent vector perturbations
[in Fourier space $ k^2 S_i(k) = 0 $]. Therefore, the \emph{vector
perturbations} are \emph{negligible} (as for classically driven
inflation \cite{multiftent}).
\be
|\delta_k^{(V)}(m^2, \lambda)|^2 = 0 \;.
\ee

\subsubsection{Tensor perturbations}

As the energy-momentum tensor of a scalar field do not have tensor
perturbations, the tensor metric perturbations [Eq. (\ref{dgT})]
do not have any source in their equation. Therefore, the amplitude
of tensor perturbations is determined only by the background
evolution, which after the condensate formation has {\it an effective
single-field classical description}. Thus, the tensor
perturbations for the quantum inflation scenario are
\be \label{tenspert}
 |\delta_k^{(T)}(m^2,\lambda)|^2 =
    \left| \tilde\delta_k^{(T)}\left(m^2,\;\frac{\lambda}{2 N} \right)
   \right|^2
\ee
where $ |\tilde\delta_k^{(T)}(\tilde m^2,\tilde{\lambda})|^2 $ is
the power spectrum of tensor perturbations for {\it single}-field
classical inflation.

It can be expressed in terms of the slow roll parameters
[see Appendix] as
\bea\label{dT}
|\tilde\delta_k^{(T)}(\tilde m^2,\tilde{\lambda})|^2 &=&
  \frac{2}{\pi^2 \;  M_{Pl}^2} \; H_{\cal H}^2 \;
       [1+2(1-\gamma-\ln 2) \;   \epsilon_{\cal H}] \cr \cr 
 &=& \frac23 \; \frac{V}{\pi^2 \;  M_{Pl}^4}\left[ 1 + \left(\frac76 -
	 \gamma-\ln 2 \right)  M_{Pl}^2 \; \left(
	 \frac{V'}{V} \right)^2 \right] \quad .
\eea
Using the relation Eq.(\ref{tenspert}) we can compute the
tensor spectral index from the single-field classical inflation
result, which in terms of the slow roll parameters is:
\bea\label{dT2}
&&n_T = - 2 \;  \epsilon \; , \cr \cr 
&& \frac{d n_T}{d \ln k} = 4 \; \epsilon \; \eta_V - 8 \; \epsilon^2 = -
n_T\left( n_S - 1 - n_T \right) \; , \cr \cr 
&& n_T = -  M_{Pl}^2 \; \left(\frac{V'}{V} \right)^2\; , 
\label{nt} \cr \cr 
&& \frac{d n_T}{d \ln k} = -2 \;  M_{Pl}^4  \left(\frac{V'}{V}\right)^2
\left[ \left(\frac{V'}{V}\right)^2 - \frac{V''}{V}  \right] \; .
\eea
taken at the value of the field when the scale of interest exited the 
horizon.

The tensor perturbations (\ref{dT}) and (\ref{dT2})
only get quantum inflaton corrections from the changes in the
metric background. These corrections, induced by $ \delta{\cal M}^2 $ 
[Eqs. (\ref{delL})-(\ref{delN})] through the Friedmann equation, are
of the order $1/N$. They could be similar to those appearing in classical 
inflation since they arise from a change of the metric background.

For example, for chaotic inflation in the limiting cases $
\tilde\varphi \ll m / \sqrt{\lambda} $ and $
\tilde\varphi \gg  m / \sqrt{\lambda} $,
 the potential and the inflaton field are  given by Eqs.(\ref{potb}) and
(\ref{camf}), respectively. Plugging these results in
Eq.(\ref{nt}) yields,
\be \label{nTb}
n_T = - \frac{b}{2N_e} \quad ,  \quad\frac{d n_T}{d \ln k} = -
\frac{b}{2 \, N_e^2} \;.
\ee
Evaluating this expression at $ N_e = 60 $ efolds before the end of
inflation, if the quadratic term dominates we obtain
\be
n_T = - 0.017  \quad ,  \quad\frac{d n_T}{d \ln k} = -0.0003 \;;
\ee
while if the quartic term dominates we have
\be
n_T = - 0.033  \quad ,  \quad\frac{d n_T}{d \ln k} = -0.0005\;.
\ee
Future measurements for the amplitude and  spectral index of
tensor perturbations are  important because the relation
between $ n_T $ and the tensor to scalar amplitude ratio is model
dependent, and therefore it will allow to discriminate between
inflationary models (see below).

\subsubsection{Tensor to scalar amplitude ratio}

The tensor to scalar ratio $ r $ is defined as
\be
r \equiv \frac{|\delta_{k}^{(T)}|^2}{|\delta_{k\;ad}^{(S)}|^2} \;.
\ee
From the previous expressions for the spectra of tensor [Eq.
\eqref{tenspert}] and adiabatic scalar perturbations
[Eq. \eqref{scalpert} and \eqref{tenspert}] respectively, it follows that
\be
r(m^2,\lambda) = \frac{| \delta_k^{(T)}(m^2,\lambda)|^2}{|
  \delta_{k ad}^{(S)}(m^2,\lambda)|^2}
= \frac{\left|\tilde\delta_k^{(T)}\left(m^2,\;\frac{\lambda}{2 N}
   \right) \right|^2}
   {\left| \tilde\delta_{k\;ad}^{(S)}\left(m^2,\;\frac{\lambda}{2 N} \right)
   \right|^2}
= \tilde r \left(m^2,\;\frac{\lambda}{2 N} \right)
\ee
where $ \tilde r (m^2,\lambda) $ is the tensor
to scalar amplitude ratio for single-field classical inflation.

Expressing this result in terms of the slow roll parameters
Eqs.(\ref{epseta}) we obtain 
\bea\label{rT}
r &=& 16 \; \epsilon_{\cal H}  \left[ 1 - 2 \; 
\left(2 - \gamma-\ln 2 \right)\;
 \left( \epsilon_{\cal H} + \eta_{\cal H} \right) \right] + 
{\cal O}\left(
\epsilon_{\cal H}^2, \eta_{\cal H} \; \epsilon_{\cal H}, \eta_{\cal H}^2 
 \right) \cr \cr
 & = & 8 \;  M_{Pl}^2 \; \left(\frac{V'}{V} \right)^2 \left[ 1 - 2 \;  
\left(2 - \gamma-\ln 2 \right)  M_{Pl}^2 \; \frac{V''}{V} \right] \; ,
\eea
and the following consistency relation at leading order
\be \label{rnT}
 n_{T} = - \frac{r}{8}  \quad ,  \quad\frac{d n_T}{d \ln k}
 =\frac{r}{8} \left[ n_S - 1 +\frac{r}{8} \right] \; ,
\ee
which is the same as for single-field classical inflation 
\cite{multiftent}. 

This consistency relation is model dependent, 
therefore simultaneous measurement of $ r $ and $ n_{T} $
will select between inflationary models. In particular the consistency
relation Eq.(\ref{rnT}) will change when isocurvature scalar 
perturbations are present. 

\medskip

For example, for chaotic inflation in the limiting cases $
\tilde\varphi \ll  m / \sqrt{\lambda} $ and $
\tilde\varphi \gg  m / \sqrt{\lambda} $,
 the potential and the field are  given by Eqs. (\ref{potb}) and
(\ref{camf}), respectively. Using the relation \eqref{rnT} and the 
result in
 Eq. \eqref{nTb} yields,
\be \label{rb}
r = \frac{8b}{2N_e}\;.
\ee
Evaluating this expression $ 60 $ efolds before the end of
inflation, if the quadratic term dominates we obtain
\be
r = 0.13  \;;
\ee
while if the quartic term dominates we have
\be
r = 0.27  \;.
\ee
From the WMAP$_{ext}+2$dFGRS+Lyman $\alpha$  data\cite{wmap1,wmap2,wmap3}, 
the upper bound for $ r $ is 
$$
r(k_0=0.002\ {\rm Mpc}^{-1})<0.90 \quad (95\% CL) \; . 
$$
The maximum likelihood single-field inflationary model for the 
WMAP$_{ext} + 2$dFGRS+Lyman $ \alpha $ data set has $ r = 0.42 $. Detection 
and measurement of gravity wave power spectrum will be a further key test
for inflation. 

The no-prior $r$-limit $ r < 0.90 $ along with the $2-\sigma$ upper limit
on the amplitude $ A(k_0=0.002\ {\rm Mpc}^{-1}) = 0.75^{+0.08}_{-0.09} 
(68\% CL) $, implies that the energy scale of inflation is:
$$
V^{\frac14} < 3.3 \times 10^{16} \; \mbox{GeV}
$$
at $95\%$ confidence level.

\bigskip

The two limiting examples of the quadratic and quartic potentials (for which 
$ \eta_V = \epsilon = 2 \left( \frac{M_{Pl}}{\varphi}\right)^2 $ and $ \eta_V 
= \frac32 \; \epsilon $ respectively), fall in the class B potentials models
$(0 \leq \eta_V \leq 2 \; \epsilon )$ in WMAP 
classification\cite{wmap1,wmap2,wmap3}. 
The WMAP data analysis give for this class
$$
0.94 \leq n_S \leq 1.01 \quad , \quad -0.02  \leq \frac{dn_S}{d \ln k} 
\leq 0.01 \quad , \quad 0.007  \leq r  \leq 0.26 \; .
$$
A pure monomial quartic potential (minimally  coupled) is disfavoured at
more than $3-\sigma$ by  WMAP data \cite{wmap1,wmap2,wmap3} since a too 
large $r$ is produced. 

\bigskip

We want to stress that excluding the quadratic mass term in the potential 
$ V(\varphi) $ implies a non-generic choice which is only justified at 
isolated points (critical points in statistical mechanics). 
Therefore, from a purely theoretical point of view, the pure quartic potential
is a weird choice implying to fine tune to zero the coefficient of the 
quadratic term.

As stated at the beginning of the Introduction, the inflaton field must be
considered an effective description of matter in the GUT scale in a
Ginsburg-Landau approach. Therefore, in this context the inflaton potential 
$ V(\varphi) $ should be generically a polynomial of degree four. [Higher
degree terms should be irrelevant]. Moreover, shifting the minimum of 
$ V(\varphi) $ to the origin implies that the lowest order term must
be quadratic. A cubic term cannot be excluded in general but most
potentials are assumed to be even for symmetry. This argument lefts us with
a quadratic plus quartic polynomial as in Eq.(\ref{poti}).

\section{Conclusions}  \label{conclu}

We present in this article, {\bf both} for the background and the 
perturbations, a complete quantum field treatment of inflation that
takes into account the nonperturbative quantum dynamics of the
inflaton consistently coupled to the dynamics of the 
(classical) metric. We avoid in quantum 
inflation the unnatural requirements of an
initial state with all the energy in the zero mode and breaking
the $ \vec\varphi \to -\vec\varphi $ symmetry of the potential.
For new inflation this quantum framework allows a consistent
treatment of the explosive particle production due to spinodal
instabilities.

Quantum field inflation (under conditions that are
the quantum analog of slow roll) leads, upon evolution, to the
formation of a condensate starting a regime of effective classical
field inflation. That is, the $N$-component quantum
inflaton becomes an effective $N$-component classical inflaton,
which can be directly expressed in terms of an effective
\emph{single}-field classical inflation scenario.
The action structure, parameters (mass and
coupling) and initial conditions for the effective classical field
description are fixed by those of the underlying quantum field
inflation. 

We show that this effective description allows an easy computation
of the primordial perturbations which takes into account the
dominant quantum effects (quantum inflaton background and quantum
inflaton and metric perturbations). The computation gives the
primordial perturbations for quantum field inflation in terms of
the classical inflation results. 

In particular, isocurvature
scalar perturbations are absent (at first order of slow roll) due
to the $ O(N) $ invariance of the potential  in agreement with the WMAP data.
More general non-symmetric
potentials with different masses or couplings for the different
components of the field would lead to non negligible isocurvature
density perturbations. It is thus the 
presence of a large symmetry in multi field 
models that make them compatible with the present observations.

Quantum field inflation provides enough efolds of inflation provided the 
generalized slow roll condition is fulfilled. 
In the case of chaotic quantum field inflation the number of efolds is 
lower
than in classical inflation when modes with non-zero wavenumber
 are excited initially as shown in Eq.(\ref{nefq}). As in classical 
inflation,
the primordial spectrum of perturbations turns to be 
independent of the details of the initial quantum state. 
Quantum corrections to the power spectrum turn to be approximately
of the order of $ \frac{m^2}{N \; H^2} \sim 4 \% $ for chaotic inflation.
[This value is an upper estimate corresponding to $ N = O(1) $].

In summary, the classical inflationary scenario emerges as an
effective description of the post-condensate inflationary period
both for the {\em background} and for the {\em perturbations}.
Therefore, this generalized  inflation provides the {\em quantum
field foundations} for classical inflation, which is in agreement
with CMB anisotropy observations \cite{wmap1,wmap2,wmap3,otros}.

\section*{Acknowledgments}

We thank Daniel Boyanovsky for discussions. F. J. C.
acknowledges support from MECD, UCM and MCYT (Spain) through
research projects EX-2002-0060, PR1/03-11595 and
BFM2003-02547/FISI, respectively. This work is supported in part by the 
Conseil Scientifique de l'Observatoire de Paris through an `Action Incitative'.

\appendix

\section{Scalar perturbations in multiple-field inflation}
\label{apPertMultif}

We treat here the generation of scalar density perturbations by multi field
inflatons.

The approach of ref. \cite{multiftent} is particularly useful in
the model under study. The procedure begins defining a basis in
the  internal $ \vec{\varphi} $ field space which allows a physical
interpretation of the various scalar field components. The first
basis vector, $ \vec e_1 $, is the direction of the velocity of
the field
\be
\vec e_1 =
\frac{\dot{\vec{\tilde\varphi}}}{\left|\dot{\vec{\tilde\varphi}}\right|}
\;.
\ee
where the dot stands for the derivative with respect of the cosmic time.

Next, $ \vec e_2 $ is the direction of that part of the field
acceleration $ \ddot{\vec{\tilde\varphi}} $ which is perpendicular
to $ \vec e_1 $, and this procedure is extended to higher order
derivatives to define the other basis vectors. The most important
projectors defined by this basis are
\be \label{Pdef}
{\bf P}^{\parallel} \equiv \vec e_1 \vec e_1^{ T} \;; \quad\quad
 {\bf P}^{\bot} \equiv {\bf 1} - {\bf P}^{\parallel} \;;
\ee
(superindex $ ^T $ meaning dual).

In multifield inflation the leading slow-roll functions are given by
\be\label{ante}
\epsilon \equiv - \frac{\dot H}{H^2} \;; \quad\quad
\vec{\eta} \equiv
\frac{\ddot{\vec{\tilde\varphi}}}{ H \left|\dot{\vec{\tilde\varphi}}\right|} 
\;.
\ee
The later can be decomposed in the components of $ \vec{\eta} $
parallel and perpendicular to the field velocity ($
\dot{\vec{\tilde\varphi}} $), \emph{i.e.},
\be \label{prim}
\eta^{\parallel} \equiv \vec e_1 \cdot \vec{\eta} \quad
\mbox{and}  \quad  \eta^{\bot} \equiv \vec e_2 \cdot \vec{\eta}
\ee
(by construction there are no other components).

Next, it is convenient to define the gauge invariant variable $ \vec q $,
\be
\vec q \equiv a \left( \delta\vec{\tilde\varphi} +
\frac{\Phi}{\cal H}\; \frac{d\vec{\tilde\varphi}}{d {\cal T}} \right) \; .
\ee
where $ \delta\vec{\tilde\varphi} $ is the inflaton perturbation,
$ \Phi $ is the scalar metric perturbation [Eq. \eqref{dgS}], and
$ {\cal H} = a \; H $. In terms of $ \vec q $ the evolution
equation for the perturbations in terms of the conformal time takes
the form \cite{multiftent} 
\be \label{qeq}
\frac{d^2 \vec q_{\vec k}}{d {\cal T}^2 }
 + (k^2 + {\cal H} \; {\bf \Omega}) \, \vec
q_{\vec k} = 0 \;,
\ee
where $ \vec q_{\vec k} $ is defined by
$$
\vec q = \frac{1}{(2\pi)^{3/2}} \int d^3 k \; [\vec q_{\vec k} \;
e^{-i \vec
    k \cdot \vec x} + \vec q_{\vec k}^{\,*} \; e^{-i \vec k \cdot \vec x}
] \; ,
$$
and
\be
{\bf \Omega} \equiv \frac{1}{H^2} \, \frac{\partial^2
V}{\partial\varphi^a \partial\varphi^b} - (2-\epsilon) \;  {\bf 1}
- 2 \;  \epsilon \left[ (3+\epsilon) {\bf P}^\parallel + \vec e_1
\;  \vec{\eta}^{\,T} + \vec{\eta} \;  \vec e_1^{\,T} \right]
\ee
$ \epsilon $, $ {\bf P} $ and $ \vec\eta $ are defined by Eqs.
\eqref{Pdef} and \eqref{ante}.

After quantizing the fields we can expand $ \vec{\hat q}_k $ as
\be
\vec{\hat q}_{\vec k} = {\bf Q}({\cal T}) \; \vec{\hat
c}^{\,\dag}_{\vec k} + {\bf Q}^*({\cal T}) \; \vec{\hat c}_{\vec
k} \;,
\ee
with $ \vec{\hat c}^{\,\dag}_{\vec k} $; $ \vec{\hat c}_{\vec k} $
constant creation and annihilation operator vectors and $ {\bf
Q}({\cal T}) $ a matrix function of time $ \cal T $ 
which satisfies Eq. \eqref{qeq}.
The initial conditions for the cosmological relevant modes are
vacuum initial conditions. 

Applying this procedure, the adiabatic and
isocurvature contributions to $ 
\hat\Phi_{\vec k} =
\hat\Phi_{\vec k\; ad} + \hat\Phi_{\vec k\; iso}$ at leading
order of the slow-roll approximation are
\bea
\hat\Phi_{\vec k\; ad} &=& \frac{3}{5} \;  \frac{\kappa}{2k^{3/2}} \;
  \frac{H_{\cal H}}{\sqrt{\epsilon_{\cal H}}} \; (\vec e_1^{\,T} +
  \vec U_{P \; e}^{\,T}) \; {\bf E}_{\cal H} \; \vec{\hat c}^{\,\dag}_{\vec k}
  + c.c. \label{Phiadgen}\\
\hat\Phi_{\vec k\; iso} &=& \frac{1}{6} \frac{3}{5} \;
  \frac{\kappa}{2k^{3/2}} \;  \frac{H_{\cal H}}{\sqrt{\epsilon_{\cal H}}}
  \; \vec V_e^T \; {\bf E}_{\cal H} \; \vec{\hat c}^{\,\dag}_{\vec k} + c.c.
  \label{Phiisogen}
\eea
({\em c.c.} meaning the adjoint of the previous terms)
where,
\bea\label{bodr}
{\bf E}_{\cal H} &\equiv& (1-\epsilon_{\cal H}) \; {\bf 1} +
  [2-\gamma-\ln 2] \;  {\bf \delta}_{\cal H} \;, \quad\quad
{\bf \delta} \equiv \epsilon \; {\bf 1} - \frac{1}{3H^2} \;
  \frac{\partial^2V}{\partial{\tilde \varphi^a} \partial{\tilde\varphi^b}} +
  2 \;  \epsilon \; \vec e_1  \; \vec e_1^{\,T} \;, \\
\vec U_{P \; e}^T &\equiv& 2 \sqrt{\epsilon_{\cal H}}
  \int_{t_{\cal H}}^{t_e} dt' \;  H \;  \frac{\eta^\bot}{\sqrt{\epsilon}} \;
  \frac{a_{\cal H}}{a} \; \vec e_2^{\,T} \; {\bf Q} \; {\bf Q}_{\cal H}^{-1}
  \quad ,  \cr \cr 
\vec V_e^T &\equiv& \sqrt{\epsilon_{\cal H}} \;
  \frac{\sqrt{\epsilon_e} \eta^\bot} {\epsilon_e + \eta^\parallel} \;
  \frac{a_{\cal H}}{a_e} \; \vec e_2^{\,T} \; {\bf Q}_e \,
  {\bf Q}_{\cal H}^{-1}
  \;,
\eea
with $ \gamma = 0.57721\ldots $ the Euler constant, $ \kappa^2
\equiv 8 \pi G = 8 \pi / m_{Pl}^2 $, the subscript $ e $ meaning that
the quantity has to be evaluated at the end of inflation, at a
time, $ t_e $, while the subscript $ {\cal H} $ means it has to be
evaluated at the time of horizon crossing, $ t_{\cal H} $, when $
{\cal H} = k $ (\emph{i.e.}, $ H \, a = k $) during inflation.

Scalar density perturbations in multi field inflation has been
considered in refs. \cite{multiftent, multifmuk, multifpolar}. 
In the general case, the spectra of scalar perturbations arising
from adiabatic, isocurvature and mixture of adiabatic and isocurvature
contributions, are 
\bea \label{genscalpertsr}
|\tilde{\delta}_{\vec k\; ad}^{(S)}|^2 &\equiv&
  \frac{2k^3}{9\pi^2}\; \langle \hat\Phi_{\vec k\; ad}^2 \rangle
=
  \frac{\kappa^2}{50\pi^2} \;  \frac{H_{\cal H}^2}{\epsilon_{\cal H}} \;
  \left[ (1-2 \; \epsilon_{\cal H})
  (1 + \vec U_{P \; e}^T \vec U_{P \; e})+ \right. \cr\cr
&& \left. + 2 (2-\gamma-\ln 2)\; (2 \; \epsilon_{\cal H}
  + \eta_{\cal H}^\parallel + 2\; \eta_{\cal H}^\bot e_2^T \vec U_{P\; e}
  + \vec U_{P \; e}^T \delta_{\cal H} \vec U_{P \; e}) \right]
  \;.  \nonumber \\
| \tilde\delta_{k\; mix}^{(S)}|^2 &\equiv&
  \frac{2 k^3}{9\pi^2} \;  (\langle \hat\Phi_{k\; iso} \hat\Phi_{k\; ad}
  \rangle + \langle \hat\Phi_{k\; ad} \hat\Phi_{k\; iso}
  \rangle ) = \cr\cr
&& = \frac{1}{6} \frac{\kappa^2}{50\pi^2}
  \frac{H_{\cal H}^2}{\epsilon_{\cal H}}
  \left[ (1-2\epsilon_{\cal H}) \vec U_{P \; e}^T V_e
  + 2 (2-\gamma-\ln 2)\; (\eta_{\cal H}^\bot e_2^T V_e
  + \vec U_{P \; e}^T \delta_{\cal H} V_e) \right] \\
| \tilde\delta_{k\; iso}^{(S)}|^2 &\equiv&
  \frac{2 k^3}{9\pi^2} \;  \langle \hat\Phi_{k\; iso}^2 \rangle
=
  \frac{1}{36} \frac{\kappa^2}{50\pi^2}
  \frac{H_{\cal H}^2}{\epsilon_{\cal H}}
  \left[ (1-2\epsilon_{\cal H}) V_e^T V_e
  + 2(2-\gamma-\ln 2)\; V_e^T \delta_{\cal H} V_e \right] \nonumber
\eea
Coupling between perturbations in the different field components generate
entropy or isocurvature perturbations.

A generic multi field model has all three contributions of the same
order. However, if the potential is completely symmetric for $ O(N) $
rotations in the internal space, the slow roll trajectories are
straight lines in field space. We show below that in this case
the isocurvature density perturbations are negligible and the adiabatic 
contributions dominate. The adiabatic density perturbations are then the same 
as in the {\it single-field case}. 

\subsection{Scalar perturbations for a straight trajectory 
 of the background field}

We particularize in this section the previous results to the case
when the trajectory of the background field is a straight
trajectory in the internal field space. In this case $ \vec \eta $
Eqs.(\ref{ante}) and (\ref{prim}) has,
\be
\eta^\bot = 0 \quad , \quad \eta^\parallel \equiv \eta 
\ee
that implies
\be\label{UV}
\vec U_{P \; e}^T = 0 \quad ; \quad \quad
\vec V_e^T = 0 \;,
\ee
using Eq. \eqref{Phiadgen}-\eqref{Phiisogen},
\bea\label{FHE}
\hat\Phi_{\vec k\; ad} &=& \frac{3}{5} \frac{\kappa}{2k^{3/2}} \;
  \frac{H_{\cal H}}{\sqrt{\epsilon_{\cal H}}} \; \vec e_1^{\,T}
  \; {\bf E}_{\cal H} \; \vec{\hat c}^{\,\dag}_{\vec k} + c.c. \label{Phiad}\\
\hat\Phi_{\vec k\; iso} &=& 0 \;.
\eea
Therefore, the scalar isocurvature perturbations are negligible at
leading order of slow-roll when the bulk inflaton trajectory is a
straight line, and we have
\bea\label{bodri}
| \tilde\delta_{k\; iso}^{(S)}|^2 &\equiv&
  \frac{2 k^3}{9\pi^2} \;  \langle \hat\Phi_{k\; iso}^2
  \rangle = 0 \;,\\ 
| \tilde\delta_{k\; mix}^{(S)}|^2 &\equiv&
  \frac{2 k^3}{9\pi^2} \;  (\langle \hat\Phi_{k\; iso} \hat\Phi_{k\; ad}
  \rangle + \langle \hat\Phi_{k\; ad} \hat\Phi_{k\; iso}
  \rangle ) = 0 \;.
\eea
From Eqs.(\ref{bodr}), (\ref{UV}) and (\ref{FHE}),
the scalar adiabatic perturbations are 
\be \label{adscalpertsr}
|\tilde{\delta}_{\vec k\; ad}^{(S)}|^2 \equiv \frac{25 k^3}{18\pi^2}
\; \langle \hat\Phi_{\vec k\; ad}^2 \rangle =
  \frac{1}{8 \, \pi^2 \, M_{Pl}^2} \;  
\frac{H_{\cal H}^2}{\epsilon_{\cal H}} \;
  \left\{ 1-2 \; \epsilon_{\cal H} + 2 [2-\gamma-\ln 2]\;
  (2 \; \epsilon_{\cal H} + \eta_{\cal H}) 
+ {\cal O} \left(\epsilon_{\cal H}^2,
\eta_{\cal H}^2,\eta_{\cal H}\epsilon_{\cal H}\right)
  \right\} \;.
\ee
\emph{i.e.}, the adiabatic perturbations are the same as for a
single-field inflation which only
takes into account the field component in the direction of $ \vec
e_1 $. (Recall that the subscript $ {\cal H} $ means that the
quantity has to be evaluated at the time of horizon crossing, when
$ H \, a = k $.) We used here the same normalisation convention as the WMAP 
collaboration \cite{wmap1,wmap2,wmap3}.
$ |\tilde{\delta}_{k\;ad}^{(S)}(\tilde m^2,\tilde{\lambda})|^2 $in 
Eq.(\ref{scalpertsr}) corresponds to the WMAP amplitude 
$ \Delta_{\cal R}^2(k) $. WMAP only uses the leading order in slow roll.

\section{Tensor perturbations in multiple-field inflation}
\label{aptenper}

In the general multi field inflaton case, the energy-momentum of
the inflaton does not have tensor perturbations, therefore the  equation for
the tensor metric perturbations do not have any source\cite{multiftent}. 
Thus, the tensor perturbations are determined only by the background evolution.

The Einstein equations for the tensor perturbations are:
\be
h_{ij}'' + 2 {\cal H} h_{ij}' - \Delta h_{ij} = 0 \;.
\ee
$ h_{ij} $ is symmetric, transverse and traceless. Thus, each of
its Fourier modes only has two independent components, and they
can be decomposed as
\be
h_{ij}(\vec k) = h^+(\vec k) \, e_{ij}^+(\vec k) + h^\times(\vec
k) \, e_{ij}^\times(\vec k) \;,
\ee
where $ e_{ij}^+ $ and $ e_{ij}^\times $ are the polarization
tensors. (In a coordinate system where $ \vec k $ points along the
$z$-axis, the nonzero components are $ e_{xx}^+ = - e_{yy}^+ = 1 $
and $ e_{xy}^\times = e_{yx}^\times = 1 $.)

$ h_{ij}(\vec k) $ can be quantized and expressed as
\be
h_{ij}(\vec k) = \sum_{A=+,\times} \frac{\sqrt{2}\kappa}{a}  \,
 e_{ij}^A (\vec k) \left[ \psi_A (\vec k) \, a^\dagger_{A \vec k} +
 c.c. \right] \;,
\ee
with the creation and annihilation operators satisfying the
relations,
\be
[a_{A \vec k}, a^\dagger_{B \vec k'}] = \delta_{AB} \; \delta(\vec k
- \vec k') \;.
\ee
The equation of motion for the mode functions $ \psi_A (\vec k) $ is,
\be
\frac{d^2 \psi_{A \vec k}}{d{\cal T}^2} + \left( k^2 - \frac1a \; 
\frac{d^2 a}{d{\cal T}^2} \right) \psi_{A \vec k} = 0 \;.
\ee
After imposing vacuum initial conditions for these modes at large $
k $, the previous equations give \cite{multiftent}
\be
h_{ij} (\vec k) = \frac{\kappa}{k^{3/2}} \;  H_{\cal H} \;  [1+
\epsilon_{\cal H} \; (1-\gamma-\ln 2)] \sum_{A=+,\times} e_{ij}^A \,
\, a^\dagger_{A \vec k} + c.c. \;.
\ee
Therefore, the spectrum of tensor perturbations for multi field
inflation results
\be
|\tilde\delta_k^{(T)}(\tilde m^2,\tilde{\lambda})|^2 = \left( \frac94 \right)
\frac{2 \;
k^3}{9 \, \pi^2} \;  \langle h_{ij}(k)\, h^{ij}(k) \rangle =
  \frac{2 \; H_{\cal H}^2}{\pi^2 \; M_{Pl}^2 } \; [1+2 \; (1-\gamma-\ln 2) \; 
  \epsilon_{\cal H}] \; ,
\ee
where the factor $ \frac94 $ corresponds to the WMAP normalization convention.

\end{document}